\documentclass[12pt]{article}
\RequirePackage[OT1]{fontenc}
\RequirePackage{amsthm,amsmath}
\usepackage{breqn}
\usepackage{soul}
\usepackage{subfig}
\usepackage[countmax]{subfloat}
\usepackage[numbers,sort&compress]{natbib}
\usepackage{hyperref}
\usepackage{amsfonts}
\usepackage{amssymb}
\usepackage{multirow}
\usepackage{setspace}
\newtheorem{theorem}{Theorem}[section]

\newtheorem{problem}{Problem}[section]
\usepackage[countmax]{subfloat}
\usepackage[numbers,sort&compress]{natbib}

\DeclareGraphicsExtensions{.jpg,.pdf,.png,.jpg,.eps}

\newcommand{\blambda}{{\boldsymbol \lambda}}

\def\by{{\boldsymbol y}}

\def\b1{{\mathbf 1}}
\def\b0{{\mathbf 0}}

\def\bA{{\boldsymbol A}}

\def\bC{{\boldsymbol C}}
\def\bu{{\boldsymbol u}}

\def\bp{{\boldsymbol p}}
\def\bq{{\boldsymbol q}}

\def\bx{{\boldsymbol x}}

\def\bI{{\boldsymbol I}}
\def\bP{{\boldsymbol P}}
\def\bQ{{\boldsymbol Q}}

\def\cK{{\mathcal K}}

\def\bbR{{\mathbb R}}

\def\bbI{{\mathbb I}}

\def\btau{{\boldsymbol \tau}}
\def\blambda{{\boldsymbol \lambda}}

\begin{document}

\title{Determining a credit transition matrix from\\
cumulative default probabilities.\\
An entropy minimization approach}
\author{Henryk Gzyl$^1$ and Silvia Mayoral$^2$\\
Centro de Finanzas IESA, Caracas, Venezuela.\\
ORCID: 0000-0002-3781-8848\\
$^2$Dept. of Business Administration, Univ. Carlos III de Madrid.\\
ORCID: 0000-0002-6735-2604\\
 henryk.gzyl@iesa.edu.ve; smayoral@emp.uc3m.es}
\date{}
 \maketitle

\setlength{\textwidth}{4in}

\vskip 1 truecm
\baselineskip=1.5 \baselineskip \setlength{\textwidth}{6in}
\begin{abstract}
To quantify the changes in the credit rating of a bond is an important mathematical problem for the credit rating industry. To think of the credit rating as the state a Markov chain is an interesting proposal leading to challenges in mathematical modeling. Since cumulative default rates are more readily measurable than credit migrations, a natural question is whether the credit transition matrix (CTM) can be determined from the  knowledge of the cumulative default probabilities.

Here we use a connection between the CTM and the cumulative default probabilities to setup an ill-posed, linear inverse problem with box constraints, which we solve by an entropy minimization procedure. This approach is interesting on several counts. On the one hand, we may have less data that unknowns, and on the other hand, even when we have as much data as unknowns, the matrix connecting them may not be invertible, which makes the problem ill-posed. 

Besides developing the tools to solve the problem, we apply it to several test cases to check the performance of the method. The results are quite satisfactory.
\end{abstract}


\begin{spacing}{0.5}
\small{\tableofcontents}
\end{spacing}

\section{Introduction and Preliminaries} 
Modeling credit risk is a complicated affair. The current proposal consists of grouping obligors into categories, and provide the relevant statistical properties of each category. This leads to indices or scores, like for example the Altman Z-score, or the credit default tables provided by the rating agencies. These provide a probability of default for each of the classes as function of time. The time unit is usually taken to be a year, even though individual obligors can default any time within a given year. 

It has been noted that the risk rating of individual creditors can improve or worsen over time. For modeling purposes, it is usually agreed that once an obligor defaults, it ceases to take part in the debt market, and its risk rating stays the same (i.e., defaulted) thereafter. 

Besides assigning each debt issuer into a credit rating class, it is important to know how the credit rating evolves in time. The simplest proposal, that incorporates randomness and statistical predictability, consists of supposing that the time evolution of the rating classes is described by a time homogeneous Markov chain with one absorbing state. To suppose that the time evolution is time-homogeneous and Markovian, is a strong assumption, but it certainly provides a framework in which to analyze the problem of the time evolution of the credit rating. It provides us with a descriptive tool that allows us to answer questions like what is the mean time to default, for how long in the average a credit rating might improve or worsen. That framework also provides tools for pricing credit rating derivatives that depend on the time evolution of the risk quality of a bond issuer. 

It is nevertheless important to stress, that our modeling is carried out in discrete time. This presupposes the choice of a time unit and that default is declared at the end of a time interval. This assumption allows the use of discrete time absorbing Markov chains as the mathematical model behind the CTM methodology.

To finish this preamble, we mention that the mathematical treatment of absorbing Markov chains is a textbook matter. See Kemeny and Snell (1960), Neuts (1981) or Karlin and Pinsky (2011) for example. For connections to economics see von Neumann (1945-1946), Karlin (1959) and Petterson and Olinick (1982). For applications in biology consider Watterson (1961) and Gosselin (2001), to polymers, Mazur (1964); in animal sciences, Maw et al. (2021); in supply chain networks, Perera et al. (2019); and in cybersecurity  Subil and Suku (2014). 

Next we mention some papers on the theoretical and practical implementation of a statistical methods to determine credit transition matrices. These determine frequency of transitions between credit classes. Besides the issues related to the occurrence of the actual class transition, they also deal with the issues related to the use of continuous time modeling. See Lando and Sk\o deberg (2002), Jafry and Shuermann (2004); Schmid (2004), Jones (2005), Deng et al. (2007) consider the special case of the farm market, Tr\"{u}ck (2008), Shuermann (2008), Fei et al. (2012), Grzybowska el al. (2012), Gavalas and Syriopoulos (2014), Boreiko et al. (2018) and Ferretti et al. (2019). These works present not only different approaches to extracting information from migration data in continuous time, but also consider comparisons of the results obtained. 

In contrast to the statistical methodologies, the method that we propose, requires only the cumulative default probability as provided by the credit rating agencies plus a relationship between them and the CTM.

The remainder of this work is organized as follows. In Section 2 we invoke some well lnown properties of absorbing Markov chains to establish a connection between the credit transition matrix and the cumulative probabilities of default. After that we explain how to turn that relationship into a linear ill-posed problem for the credit transition matrix. The convex constraints on the problem come from the fact that the entries of the matrix are probabilities, and a priori fall in the interval $[0,1].$ But the method that we propose to find them allows for further specification of the range, compatible with the data provided by the rating agencies.

In Section 3 we explain the method of solution of the inverse problem minimizing an entropy of the Fermi-Dirac type, and obtain the representation of the solution in terms of the Lagrange multipliers. This is a rather interesting method that automatically yields a solution satisfying the convex constraints. 
In Section 4 we work out some examples. The numerical examples will be of two types. One in which we generate the data from a CTM provided in Lando and Sk\o deberg (2002), which serves as a test of the performance of our proposal, and another example in which we reconstruct from publicly available cumulative default probabilities. We find that our proposal satisfactorily yields the cumulative default probabilities of the years beyond the data used to infer the CTM, which in itsef is a nice consistency result.

\section{Establishing a realtionship between the CTM and the cumulative default probabilities}
In order to state the inverse problem to be solved, we need to introduce some notations. Consider a time homogeneous, discrete absorbing Markov chain $X,$ with state space $\cK=\{1,2,...,K\}.$ The states are interpreted as credit classes or credit ratings, and the state labeled by $K$ to be the default state. We write the transition matrix $\bP$ of the chain as:
\begin{equation}\label{mat1}
\bP= \begin{pmatrix} \bQ & \bp(1)\\
                      \b0^t & 1\end{pmatrix}
\end{equation}

As customary, the elements $\{P_{i,j}=P(j|i): 1\leq i,j\leq K\}$ of the transition matrix $\bP$ describe the conditional probabilities $P(X(n)=j|X(n-1)=i),$ where $X(n)$ describes the state of the system (a possible credit risk class) at time $n\geq 0.$ We reserve the notation $\bQ_{i,j}$ with $1\leq i,j\leq K-1$ for the credit transition matrix which we want to determine. Here $\b0$ stands for the $(K-1)-$zero vector, and the superscript ``$t$'' denotes the transpose of the corresponding object. Below we write $\bu$ for the $K-1$ vector with all components equal to $1.$ The $n-$step transition probability matrix is:

\begin{equation}\label{matn}
\begin{aligned}
\bP^n &= \begin{pmatrix} \bQ^n & \bp(n)\\
                      \b0^t & 1\end{pmatrix}\\
\bp(n) &= \sum_{k=1}^n\bQ^{k-1}\bp(1)=\bp(1)+\bQ\bp(n-1).
\end{aligned}
\end{equation}

The important remark here is that $(\bp(n))_i=P(X_n=K|X_0=i)=P(T\leq n|X_0=i).$ This can also be verified by invoking a first-step analysis argument, see Kemeny and Snell (1960), Neuts (1981) or Karlin and Pinski (2011). Note that: 
$$
\begin{aligned}
P(T\leq n|X_0=i)=P(T=1|X_0=i)+P(2\leq T\leq n|X_0=i)\\
 = P(T=1|X_0=i)+\sum_{j=1}^{K-1}Q(i,j)P(1\leq T\leq n-1|X_0=j)
\end{aligned}
$$
after an application of the strong Markov property in the second term in the middle equation. Thus, both $\bp(n)$ and $P(T\leq n|X_0=1)$ satisfy the same recurrence equation with the same initial condition, therefore, they coincide. The cumulative probability of default is provided by the rating agencies. An example is presented in Tables \ref{pow1} or \ref{tab3} below, which is the data that we will use in our numerical problem.

So, finally, the problem that we address here is: Determine the matrix $\bQ$ from the knowledge of the vector $\bp(n)$ for $n=1,...,N.$

To establish a system of equations from which to determine $\bQ,$ let us first introduce some more notations. The normalization condition for $P(i,j)$ is $1=\sum_{j=1}^K P(X_1=j|X_0=i),$ which after separating the last term yields the normalization condition for $\bQ$:

\begin{equation}\label{norm}
\bQ\bu=\bu^t-\bp(1)
\end{equation}
From \eqref{matn} we also have

\begin{equation}\label{norm2}
\bQ\bp(n) = \bp(n+1)-\bp(1).
\end{equation}

 Let us now introduce some more symbols.

\begin{equation}\label{data}
\begin{aligned}
&\bq(1) = \bu^t-\bp(1)\\
&\bq(n+1) = \bp(n+1)-\bp(1),\;\;\;n=1,...,N.
\end{aligned}
\end{equation}

The inverse problem consists of finding the sub-stochastic transition matrix $\bQ$ from the knowledge of the vectors $\bq(k)$ for $k=1,...,N,$ which contains the information about the consecutive credit default probabilities for a few consecutive years.  In symbols:

\begin{gather}
\mbox{Determine the}\;\; (K-1)^2 \;\;\mbox{matrix}\;\;\; \bQ\;\;\;\mbox{such that}\nonumber\\
\bQ\bu = \bq(1),\label{prob1a}\\
\bQ\bp(n) = \bq(n+1)\,\,\,n=1,...,N-1. \label{prob1b}
\end{gather}

This certainly is an ill-posed linear problem with convex constraints upon the unknown matrix $\bQ.$ 

The next step consists of vectorizing the problem into the form $\bA\bx=\by,$ where the matrix $\bA$ takes care of the constraints and, $\bx$ will stand for the vector in $\bbR^{(K-1)^2}$ obtained by listing the components of the transpose of $\bQ_{i,j}$ lexicographically. Or, if you prefer, by stacking the rows of the transition matrix $Q_{i,j}.$ Consider now the following $(K-1)\times(K-1)^2-$matrices:

\begin{equation}\label{cmat}              
\bC_0 = \begin{pmatrix}                
         \bu^t & 0 & ... & 0\\ 
           0   & \bu^t &...& 0\\
	  \vdots&\vdots&...&\vdots\\
            0  &  0  &... & \bu^t
\end{pmatrix},\;\;
\bC_k = \begin{pmatrix}                
         \bp^t(k) & 0 & ... & 0\\ 
           0   & \bp^t(k) &...& 0\\
	  \vdots&\vdots&...&0\\
            0  &  0  &... & \bp^t(k)
\end{pmatrix}, \;\;\;k=1,...,N-1.
\end{equation}
To finish, the matrix $N(K-1)\times(K-1)^2$ matrix $\bA$ and the $N(K-1)-$dimensional data vector $\by$ are defined to be:

\begin{equation}\label{data1}
\bA  = \begin{pmatrix}
          \bC_0\\
         \bC(1)\\
           \vdots\\
         \bC(N-1)
\end{pmatrix},\;\;\;
\by = \begin{pmatrix}
          \bq(1)\\
         \bq(2)\\
           \vdots\\
         \bq(N)
\end{pmatrix}.
\end{equation}
With these notations, we shall solve the following two problems: First the case with minimal box constraints 

\begin{problem}\label{prob1}
Determine $\bx\in\bbR^{(K-1)^2}$ such that $0 \leq x_j \leq 1;\,j=1,...,K^2,$ satisfying:
\begin{equation}\label{prob1.1}
\bA\bx=\by.
\end{equation}
where the $N(K-1)\times(K-1)^2$ matrix  $\bA,$ and the $N(K-1)-$vector $\by$ are specified in \eqref{data1}.
\end{problem}

\begin{problem}\label{prob2}
Determine $\bx\in\bbR^{(K-1)^2}$ such that $a_j\leq x_j\leq b_j;\,j=1,...,K^2,$ satisfying:
\begin{equation}\label{prob2.1}
\bA\bx=\by
\end{equation} 
The constraints $a_j<b_j$ are to be supplied by the analyst according to the situation. The specification amounts to preassigning a range for the unknown probabilities. In our numerical examples, we shall impose such constraints upon the diagonal elements of $\bQ.$ This knowledge is related to the observed non-default probability of each risk class. 
\end{problem}

If the number of years used to define the data vector is smaller than $K-1,$ we have an ill-posed problem in our hands. 
A glance at Table \ref{tab3} points to an additional difficulty. During the first years, there is no observed default, or no observed change in the default probability of the better rated corporations. This would cause some components of the vectors $\bq(k)$ to be zero which would force the corresponding components of the matrix $\bQ$ to be zero. When $\bQ$ is determined by counting proportions, this is taken care of by recording transitions between classes during long periods of time, see Lando and Sk\o deberg (2002), Jafry and Shuermann (2004), Jones (2005) or Shuermann (2008). 

\section{The determination of the transition matrix}

To determine the transition matrix, we first solve problem \ref{prob2} using an entropy minimization approach, and then mention how it reduces to that of problem \ref{prob1}. Let $\Psi(\bx):\prod[a_i,b_i]\to \bbR$ be defined by:

\begin{equation}\label{obj}
\Psi(\bx) = \sum_{j=1}^{(K-1)^2}\frac{x_j-a_j}{D_j}\ln\big(\frac{x_j-a_j}{D_j}\big) + \frac{b_j-x_j}{D_j}\ln\big(\frac{b_j-x_j}{D_j}\big).
\end{equation}
For problem \ref{prob1} we actually need $a_j=0$ and $b_j=1$ for all $j=1,...,K^2.$ The extra generality makes the mathematics a bit more transparent, but allows us to consider the constrained problem when need be. The objective function $\Psi(\bx)$ is strictly convex, infinitely differentiable in the interior of its domain, and its gradient is an invertible mapping in $\bbR^N.$ Its is actually a modified version of an entropy of the Fermi-Dirac type.

The Fenchel-Lagrange dual of \eqref{obj} is:

\begin{equation}\label{dual}
M(\btau) = \sum_{j=1}^{(K-1)^2}\ln\big(e^{a_j\tau_j} + e^{b_j\tau_j}\big).
\end{equation}

The interesting aspect of our choice is that \eqref{dual} is strictly convex and defined through all $\bbR^{(K-1)^2}.$ The following is a well known result from convex optimization.

\begin{theorem}\label{main}
With the notations introduced above, suppose that $\by$ is an interior point in the range of $\bA,$ and let:
\begin{equation}\label{dualent}
\Sigma(\by,\blambda)=\langle\blambda,\by\rangle - M(\bA^t\blambda).
\end{equation}
Here $\bA^t$ denotes the transpose of $\bA.$ Then the $\bx^*$ that realizes the minimum of $\Psi(\bx)$ subject to the constraints \eqref{prob2} is given by:
\begin{equation}\label{repsol1}
x_j^*= \frac{a_je^{a_j(\bA^t\blambda^*)_j} + b_je^{b_j(\bA^t\blambda^*)_j}}{e^{a_j(\bA^t\blambda^*)_j}+e^{b_j(\bA^t\blambda^t)_j}},\;\;\;\;\;\;j = 1,...(K-1)^2.
\end{equation} 
The $\blambda^*$ is to be obtained maximizing $\Sigma(\by,\blambda)$ over $\bbR^{N},$ where $N$ is the number of constraints.
Furthermore, the value of the minimal entropy is given by:
$$\Psi(\bx^*) = \Sigma(\by,\blambda^*).$$
\end{theorem}
The importance of the duality argument lies in the fact that the minimization of $\Sigma(\by,\blambda)$ is unrestricted. The only practical issue is that $\Sigma(\by,\blambda)$ may be too flat near the minimum. For that, the best thing is to combine the two-point step size gradient method, see Barzilai and Borwein (1988). The gradient of $\Sigma(\by,\blambda)$ given by:

\begin{equation}\label{grad}
\frac{\partial\Sigma}{\partial \lambda_i} = y_i - \sum_{j=1}^{(K-1)^2}A_{i,j}\frac{a_je^{a_j(\bA^t\blambda)_j} + b_je^{b_j(\bA^t\blambda)_j}}{e^{a_j(\bA^t\blambda)_j}+e^{a_j(\bA^t\blambda)_j}},\;\;\;\;\;\;i = 1,...N(K-1).
\end{equation}

The criterion for stopping the iterations is when the norm $\|\nabla_{\blambda}\Sigma\|$ given by of \eqref{grad} is smaller than some tolerance. This norm happens to be the reconstruction error, which measures how well the solutions satisfy the constraints.

To round up, once the Lagrange multipliers $\blambda^*$ are at hand, the solution to the more general problem \eqref{prob2} is given by:
\begin{equation}\label{sol2}
x_j^*= \frac{a_je^{a_j(\bA^t\blambda^*)_j} + b_je^{b_j(\bA^t\blambda^*)_j}}{e^{a_j(\bA^t\blambda^*)_j}+e^{b_j(\bA^t\blambda^t)_j}},\;\;\;\;\;\;j = 1,...(K-1)^2.
\end{equation} 

The representation of the solution to problem \ref{prob1} is obtained from that of  \eqref{sol2} by setting $a_j=0$ and $b_j=1$ for $j=1,...,(K-1)^2.$  But keep in mind that the correct Lagrange multiplier is obtained by minimizing the corresponding version of \eqref{dualent}. Anyway, the solution is given by:

\begin{equation}\label{sol1}
x_j^*= \frac{ e^{(\bA^t\blambda^*)_j}}{1+e^{(\bA^t\blambda^t)_j}},\;\;\;\;\;\;j = 1,...(K-1)^2.
\end{equation}

That \eqref{sol2} satisfies problem \ref{prob2} follows from \eqref{grad} by equating the right-hand side to $0$ which is a first order condition for $\blambda^*$ to be a maximizer of $\Sigma(\by,\blambda).$

\section{Numerical example}
In this section we develop two different examples. On the one hand, to examine the performance of the method, we use the CTM provided by Lando and Sk\o deberg (2002) as starting point, take its powers to obtain the cumulative default rates which will be the input for our procedure. Then we carry out the procedure under two types of box constraints: First with minimal constraints on all unknowns, that is $x_i\in[0,1].$ In this case we shall see that the reconstruction is good, that is, it satisfies the data and it predicts well the future default rates, but nevertheless, the discrepancy with the  known input is high. This is due to two facts. On the one hand, the problem has infinitely many solutions, and on the other hand, because the entropy minimization procedure tends to spread the unknown probabilities as much as possible.

This example emphasizes the power of our approach in handling the constraints. We only constrain the diagonal cells of the unknown matrix, and the agreement with the test solution is quite good. Note that the constraints on the diagonal cells can be obtained from the default rates of the first years. If $p_i(1)$ denotes the probability of defaulting by the end of the first year starting from credit rating $i=1,2,...,K,$ then $1-p_i(1)$ is a natural upper bound for the probability of being in the same credit class by the end of the year.

As a test of the robustness of the method, we consider information about the default rates of $K=4,5,6$ and $7$ years. To have a consistency test, we take powers of the full transition matrix to obtain predictors or forecast of the default rates, for later to compare them to the default rates provided by the entropic procedure.

For the second example, we consider the cumulative default probabilities provided in the report Macreadie (2022), which we list below. This time, we do not have an input matrix to compare with our result, but we use the reconstructions obtained from the cumulative defaults for 4,5,6,7 years to generate predictors of the cumulative default probabilities for the next 7 years of cumulative default probabilities provided in Macreadie (2022).

\subsection{Reobtaining a CTM }
As a test credit transition matrix of the underlying Markov chain we consider the example provided by Lando Sk\o deberg (2002) is:
\begin{equation}\label{TM1}
\bP = \begin{bmatrix}
    0.95912 & 0.03982 & 0.00096 & 0.00010 &  0  & 0  & 0 &  0 \\ 
    0.01249  & 0.93689 & 0.04519 & 0.00524 & 0.00015 & 0.00004 & 0 & 0\\
    0.00011 & 0.01666 & 0.93097 & 0.04906 & 0.00274 & 0.00042 & 0.00001 & 0.00003\\
    0.00002  & 0.00253 & 0.03635 & 0.90603 & 0.03955 & 0.01398 &  0.00030 & 0.00125\\
    0 & 0.00012 & 0.00318 & 0.07866 & 0.85980 & 0.05411 & 0.00317 & 0.00096\\
    0 & 0.00005 & 0.00495 & 0.00385 & 0.07029 & 0.87618 & 0.02941 & 0.01527\\
    0 & 0.00004 & 0.00091 & 0.02523 & 0.02890 & 0.11823 & 0.52289 & 0.30380\\
    0  & 0 & 0 & 0 & 0 & 0 & 0 & 1\\
\end{bmatrix}
\end{equation}
We computed the first twenty powers of $\bP$ and extract the last column of each power. Actually, it suffices to apply $\bP$ to the last column of $\bP$ twenty times. This list of cumulative default probabilities will be used to test the predictive quality of our procedure.  As data for the CMT reconstruction problem we need only the cumulative default probability for the first seven years, which we display in Table \ref{pow1}

\begin{table}[h!]
\centering
\begin{tabular}{|c|ccccccc|}\hline
   &    1   &   2     &    3    &    4     &   5      & 6      &  7   \\\hline \hline
  AAA&     0  &  0.0000  &  0.0000 & 0.0000 & 0.0000  & 0.0000   & 0.0000 \\\hline  
  AA&     0  &  0.0000  &  0.0000 & 0.0001 & 0.0001  & 0.0002  & 0.0004  \\\hline  
  A&  0.0000  & 0.0001   &  0.0003 & 0.0006  & 0.0010 & 0.0015  & 0.0022  \\\hline  
  BBB&  0.0012  &  0.0027  &  0.0045 & 0.0067  & 0.0093 & 0.0122  & 0.0154  \\\hline  
  BB&  0.0010  & 0.0037  & 0.0078  & 0.0132  & 0.0194  & 0.0263  & 0.0337  \\\hline  
  B& 0.0153  & 0.0377  & 0.0622  & 0.0865  & 0.1097  & 0.1313  & 0.1512  \\\hline  
  CCC& 0.3038  & 0.4645  & 0.5513   & 0.5998  & 0.6282 &  0.6460 &  0.6582 \\\hline
\end{tabular}
\caption{Cumulative default probabilities}
\label{pow1}			
\end{table}
Next we use this data for the unconstrained and the constrained reconstructions.

\subsubsection{Unconstrained reconstruction}
\begin{figure}[h!]
\centering
\includegraphics[width=3.0in,height=2.0in]{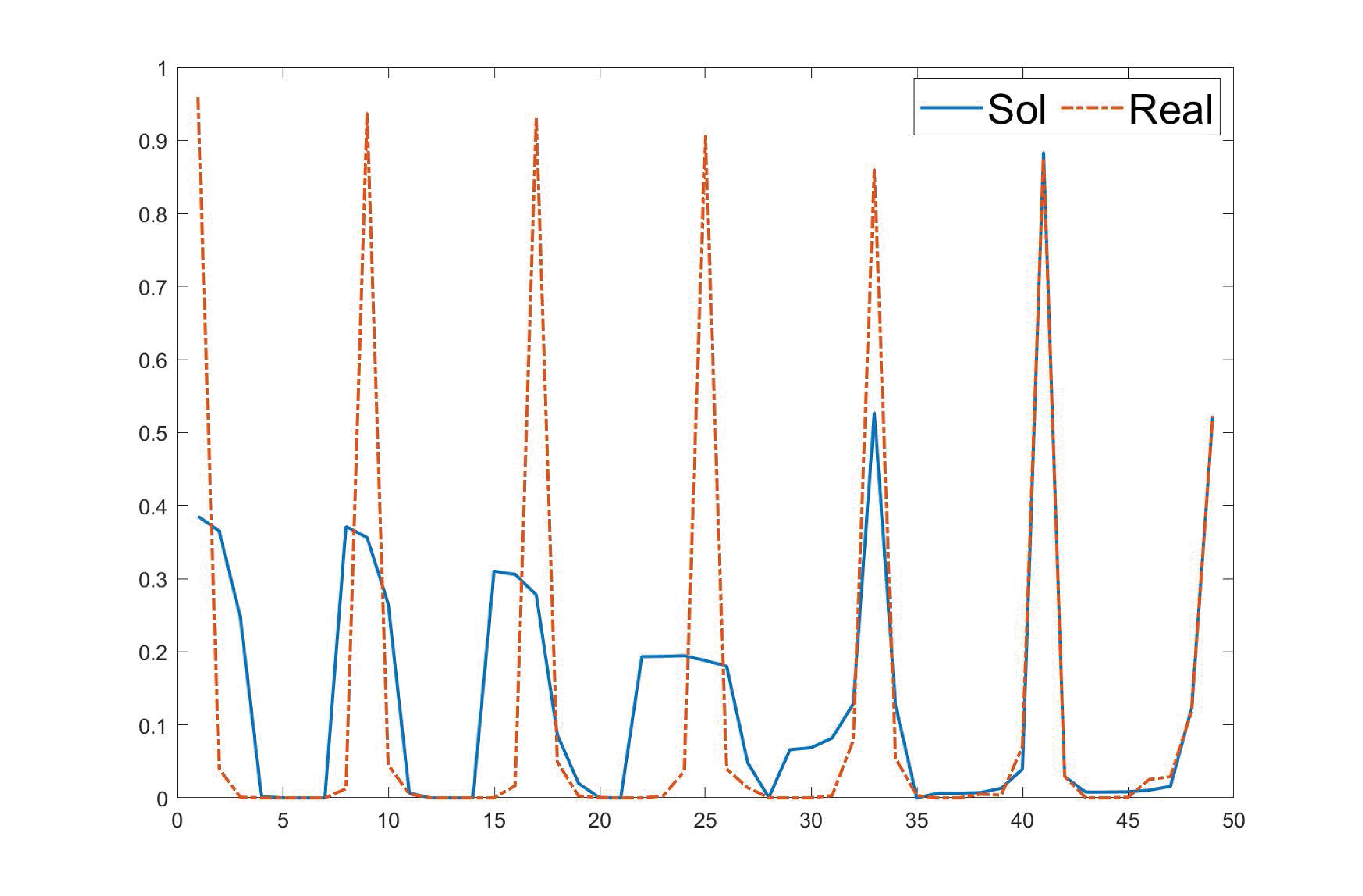}
\caption{Credit transition matrices from default data and no constraints}.
\label{CTM0} 
\end{figure}

In this case, even though the reconstruction error is of the order of $10^{-4}$ there is not much resemblance between the output and the test CTM. The result with $4,5,6,7$ years of default probabilities as data are similar. We only describe the results for the full ($7$ year) data set. The solution is given by \eqref{sol1}, and the Lagrange multiplier $\blambda^*$ is determined as explained in Section 3. Instead of listing the components of the reconstructed CTM $\bQ^*$, we plot them lexicographically (in increasing order when one moves along the successive files) in Figure \ref{CTM0} along with the components of the original test matrix $\bQ.$ The tags ``REAL'' and `SOL'' refer to the components of the test CMT and to those of the reconstructed CMT.
This test is interesting because on the one hand, the reconstruction error, given by $\|\bA\bx^*-\by\|\sim 10^{-4},$ that is, the solution provides good agreement with the data. On the other hand, even though in this case the matrix $\bA$ is square, its determinant is computed to be $\sim 10^{-150},$ which suggests the existence of infinitely many solutions.  

{\bf A consistency test: Comparing predictions}. As an additional test of the quality of the solution, we formed the matrix

\[
\bP^*= \begin{pmatrix} \bQ^* & \bp(1)\\
                      \b0^t & 1\end{pmatrix},
\]
and used it to compute the last $20$ columns of $\bP^*.$ Since we solved the problem with up to $K=7$ default probabilities, we have $13$ years of predictions to compare. We compare the cumulative probabilities predicted with the transition matrix $\bP$ against the cumulative default probabilities computed with $\bP^*$ by computing the $\ell_1$ norm of the difference of the two of them. Explicitly, if $p^*_i(n)$ (respectively $p_i(n)$) denote the $i=1,2,...,7$ component of the predicted (and respectively) the reference default probability at time $8\leq n\leq 20,$ the prediction for that year is computed as $\sum_{i=1}^7 |p_i^*(n) -p_i(n)|.$ The results are listed in Tab;e \ref{error0}.

\begin{table}[h]
\centering
\begin{tabular}{|c|ccccccc|}\hline
 Year & 8   & 9  &  10  & 11 & 12  &   13  & 14\\\hline 
      & 0.0042 &	0.0078 &	0.0125 &	0.0181 &	0.0252 &	0.0332 &	0.0421 \\\hline\hline
Year  &			15 &   16  &  17 &  18  &  19  & 20 \\\hline
      & 0.0517 &	0.0619 & 	0.0728 &	0.0841 &	0.0958 &	0.1078 & \\\hline
\end{tabular}
\caption{Error in the prediction of future cumulative default rates}
\label{error0}
\end{table}

\subsubsection{Reconstruction with box constraints along the diagonal}
For this example, as explained above, we choose the following constraints for the elements along the diagonal of $\bQ.$ We require  that the elements along the diagonal corresponding to the $AAA-AA-A$ ratings take values in $[0.9,1].$ That those corresponding to the $BBB-BB-B$ ratings take values in $[0.8,1]$ and finally we let the algorithm to assign any value in $[0,1].$ to the diagonal element corresponding to the $CCC-C$ rating. The essential constraints, in which there are zeroes in the initial default probabilities, are the first three. Therefore, $0.9$ seems a reasonable lower bound for the probability of staying in the same risk class at the end of the first year. 

\begin{figure}[h!]
\centering
\subfloat[Four years of default data]{\includegraphics[width=3.0in,height=2.0in]{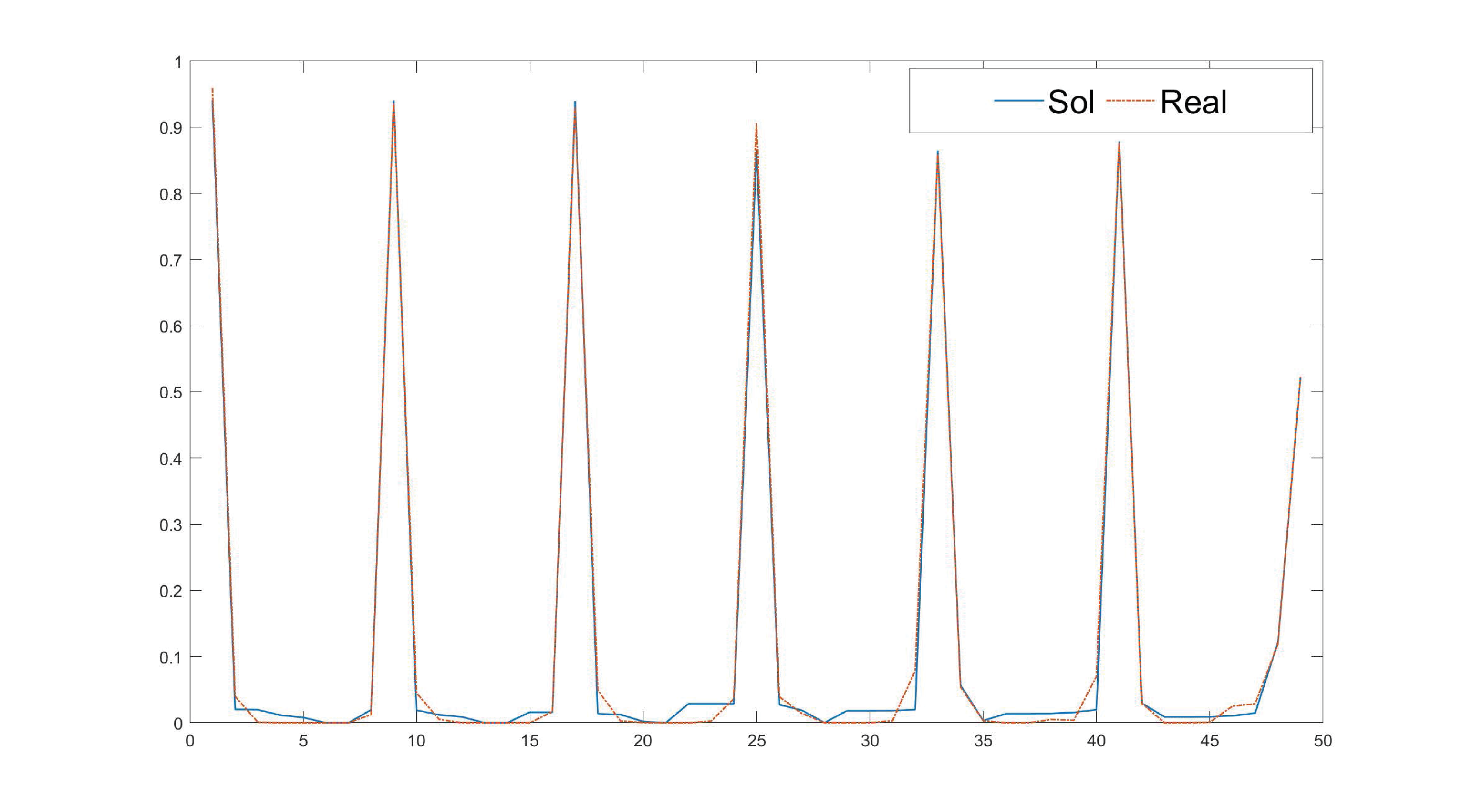}\label{P3}} 
\subfloat[Five years of default data]{\includegraphics[width=3.0in,height=2.0in]{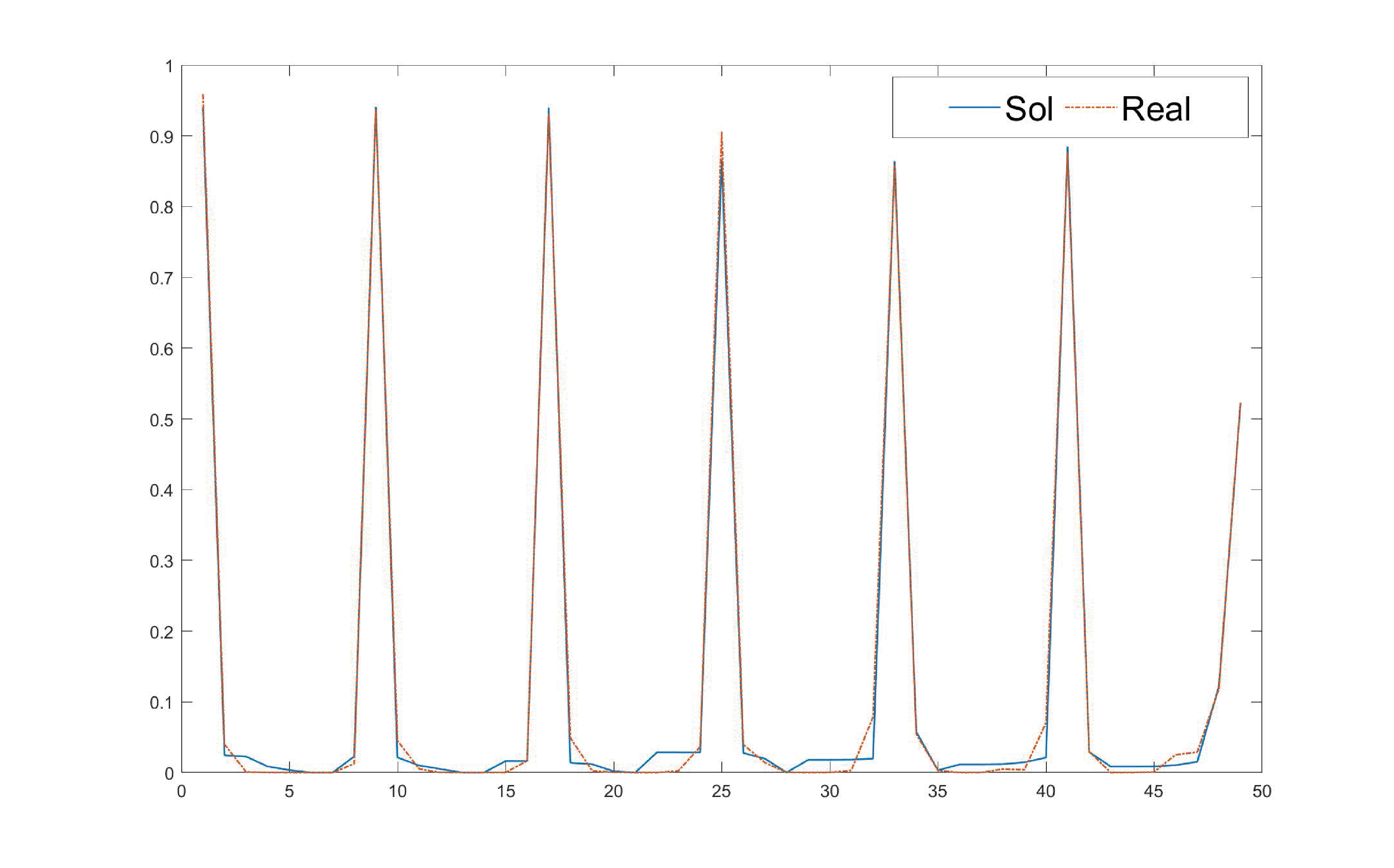}\label{P4}}

\subfloat[Six years of default data]{\includegraphics[width=3.0in,height=2.0in]{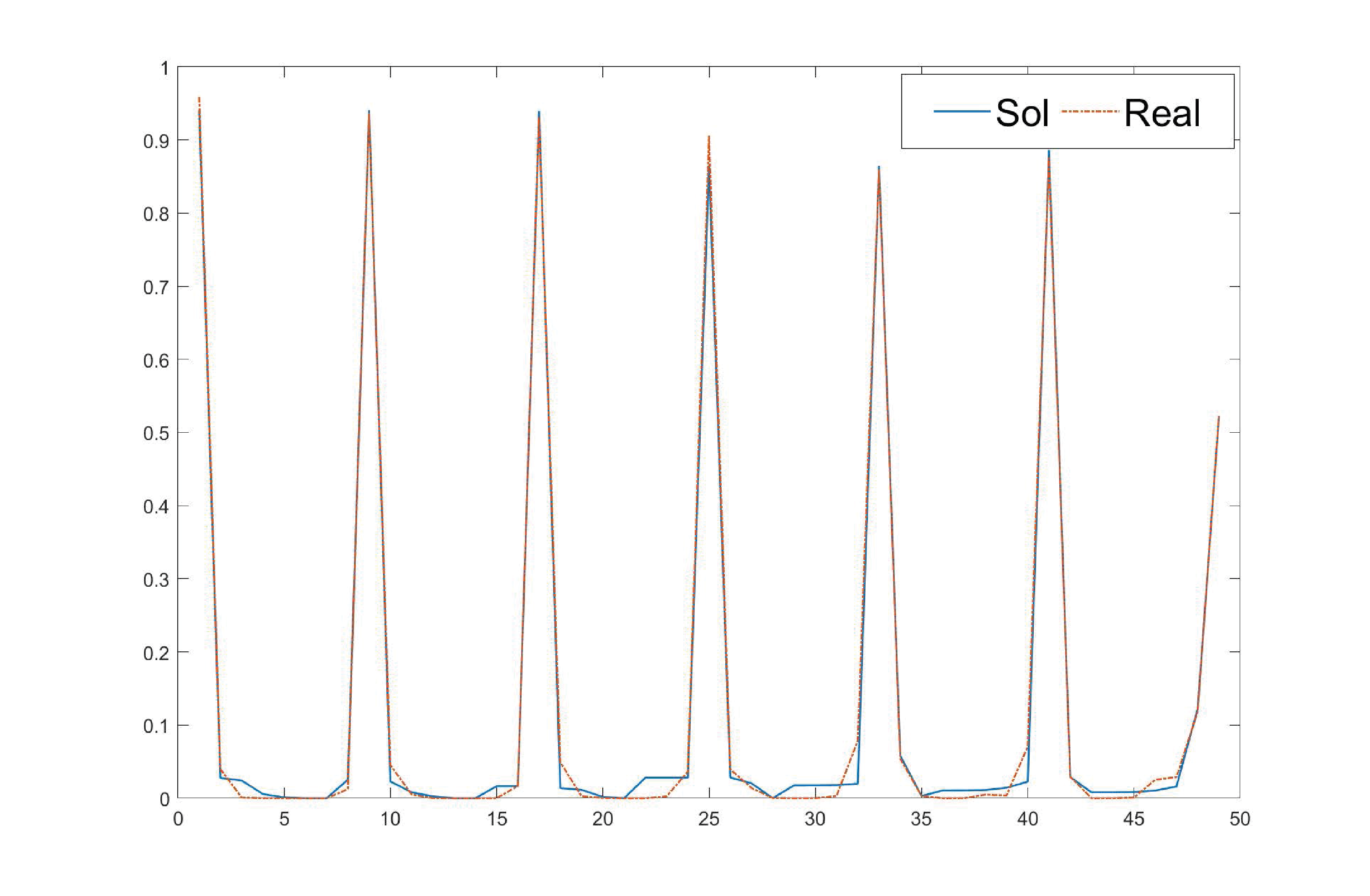}\label{P5}} 
\subfloat[Seven years of default data]{\includegraphics[width=3.0in,height=2.0in]{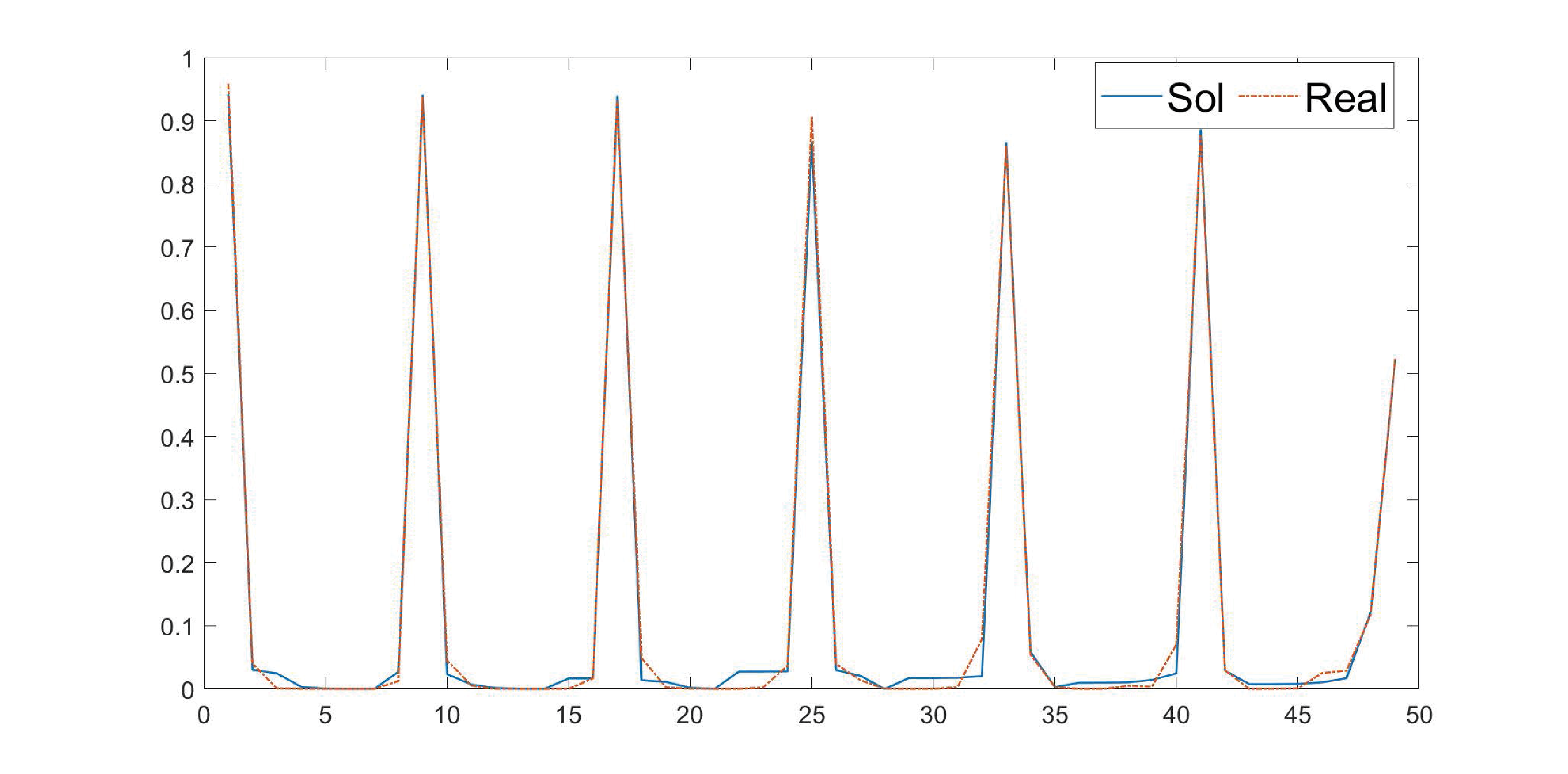}\label{P6}}
\caption{Credit transition matrices from default data}.
\label{CTM1} 
\end{figure}

The credit transition matrix $\bQ^*$ obtained by the algorithm is displayed as a plot of its vectorization along whit the plot of the original matrix The label ``REAL'' is for the test matrix,  whereas the reconstructed matrix is labeled ``SOL'' in Figure \ref{CTM1}. Recall that the vectorization of the CTM is such that the elements of the matrix are displayed in lexicographically.

The 4 panels of the figure show the CTM reconstructed, respectively, from 4,5,6 and 7 cumulative default probabilities. In each of the cases the reconstruction error, computed as the norm of the gradient \eqref{grad} was of the order of $10^{-4}.$ We also stress the fact that, when there are $7$ constraints, the matrix $\bA,$ in the statements of problem \eqref{prob1} is square, and its norm is of the order of $10^{-150}$, that is the problem of solving $\bA\bx=\by$ is as ill-posed as one can get.

{\bf A consistency text: Comparing perditions}. As a different test of the quality of the CTM $\bQ^*$ obtained above form the new transition matrix:
\[
\bP^*= \begin{pmatrix} \bQ^* & \bp(1)\\
                      \b0^t & 1\end{pmatrix}.
\]
Then we compute the first fifteen powers of $\bP^*$ and extract the last column and compare with the same last columns of the powers of the original matrix $\bP.$ In the next table we compile the $\ell_1$ norm of the difference between the two cumulative distributions as explained above.  We list only the comparison for the last eleven years of values (from year 10 to 20, the comparison of the first years is even better):

\begin{table}[h]
\centering
\footnotesize
\begin{tabular}{|c|ccccccccccc|}\hline
         &  10    &    11   &    12  &   13   &   14  &   15     &    16    &    17    &  18     &   19     &  20\\\hline
    4    & 0.0159	& 0.0209	& 0.0264 & 0.0326 &	0.0394 & 0.0407  &  0.0480  &   0.0557 & 0.0638  &  0.0723  &  0.0811\\
    5    & 0.0081	& 0.0112  &	0.0149 & 0.0192	& 0.0241 & 0.0267	 &  0.0322  &  0.0382  & 0.0446  &	0.0515  & 0.0587\\
		6    & 0.0050 & 0.0073  &	0.0102 & 0.0136 & 0.0175 & 0.0206	 &  0.0254	&   0.0306 & 0.0362	 &  0.0423	&  0.0488\\
		7    & 0.0033	 & 0.0051	& 0.0074 & 0.0102	& 0.0135 & 0.0164	 &  0.0205	&   0.0252 & 0.0304	 &  0.0362	&  0.0423 \\\hline
\end{tabular}
\normalsize
\caption{Error in the prediction of future cumulative default rates}
\label{error1}
\end{table}
In the left column of Table \ref{error1} we list the number of years of cumulative default probabilities used as data. In the top file, we list the number of years into the future that we computed the cumulative default probability. To fully describe the contents of the cells, consider, say, cell $(5,17).$ The 5 refers to the number of years of default probabilities used as data to obtain $\bQ^*$ Then we extract the last columns of $\bP^{17}$ and $\bP^{*17},$ and we form the $\ell_1$ norm of the difference of these two vectors. This is the number listed in the cell. Clearly as the number of years of data increases, the prediction becomes better, despite how small it is to begin with. We add that when the number of years into the future is smaller the predictions are better.

\subsection{Determination of the CTM from default data}
The data set for this example is the average cumulative default rates for the years 1980-2021 taken from Macreadie (2022), and displayed in Table \ref{tab3}. To choose the size of the constraints for the diagonal elements in the matrix, we do as in the last example. If as above $\bp(1)$ is the vector of default probabilities for the first year, we form $1-p_i(1)$ for $i=1,2,...,7$ and consider a number a few percentage points smaller. In the previous example we chose $0.90$ for the A-rating class, $0.80$ for the B-rating class, and $0$ for the CCC\/C rating cluster class. We repeat the same choice here.

\begin{table}[h]
\centering
\begin{tabular}{|c| c c c c c c c c|} 
 \multicolumn{9}{c}{Term(years)} \\ \hline
Rating  & 1 & 2 & 3 & 4 & 5 & 6 & 7 & 8 \\\hline
AAA     & 0.00 & 0.03 & 0.13 & 0.24 & 0.34 & 0.45 & 0.50 & 0.58 \\\hline
AA      & 0.002 & 0.06 & 0.11 & 0.20 & 0.30 & 0.40 & 0.48 & 0.55  \\\hline
A       & 0.05 & 0.13 & 0.21 & 0.32  & 0.44 & 0.57 & 0.73 & 0.87  \\\hline
BBB     & 0.15 & 0.41 & 0.72 & 1.09 & 1.48 & 1.85 & 2.18 & 2.50  \\\hline
BB      & 0.60 & 1.88 & 3.35 & 4.81 & 6.19 & 7.47 & 8.57 & 9.56 \\\hline
B       & 3.18 & 4.76 & 11.26 & 14.30 & 16.67 & 18.59 & 20.10 & 21.34  \\\hline
CCC\/C  & 26.55 & 36.74 & 41.80 & 44.74 & 46.91 & 47.95 & 49.08 & 49.82  \\\hline\hline
Rating & 9 & 10 & 11 & 12  &13  & 14 & 15 &\\\hline
AAA    & 0.64 & 0.69 & 0.72 & 0.75 & 0.78 & 0.83 & 0.89 &\\\hline 
AA     & 0.62 & 0.68 & 0.74 & 0.8 & 0.86 & 0.91 & 0.96 & \\\hline
A      & 1.01 & 1.15 & 1.28 & 1.40 & 1.52 & 1.63 & 1.76 & \\\hline
BBB    & 2.80 & 3.10 & 3.40 & 3.64 & 3.86 & 4.09 & 4.34 &\\\hline
BB     & 10.45 & 11.24 & 11.90 & 12.52 & 13.09 & 13.57 & 14.08 & \\\hline
B      & 22.45 & 23.50 & 24.40 & 25.10 & 25.75 & 26.35 & 26.92 &\\\hline
CCC\/C & 50.48 & 51.05 & 51.49 & 51.92 & 52.45 & 52.91 & 52.97 &\\\hline 
\end{tabular}
\caption{S\&P average cumulative default rates (\%), 1981-2021}
\label{tab3}
\end{table}

We ran the entropy minimization procedure with $K=4,5,6,7$  columns of the data matrix displayed in Table \ref{tab3}. In all cases the reconstruction error was of the order of $10^{-2}.$ For the case $K=7$ the matrix $\bA$ is square and has determinant $\sim 10^{-159},$ thus the problem is truly ill-posed. A possible explanation for the large reconstruction error is that the data does not necessarily come from CMT matrix. Even though there is no reference CMT matrix to compare tom S\&P has provided us with enough cumulative probability data to run a consistency test. 

{\bf The consistency test} We proceed as above. The vector $\bx^*$ is transformed into a $7\times7$ matrix $\bQ^*.$ This matrix is augmented with the first column of data plus the obvious last row to obtain the matrix $\bP^*.$ Then we compute the $15$ powers of $\bP^*$ the last column of each. The first $7$ coincide with the data, and the rest are the predictors of the cumulative defaults probabilities. We then compare to the data provided by S\&P. The comparisons are listed in Table \ref{tab4} below. There we display the differences of the $\ell_1$-norms for the given number of data points and the given number of years into the future. For example cell $(6,12)$ indicates a prediction error of about $5\%$ between the vector of cumulative default probabilities provided by S\&P and the one predicted using our methodology.

\begin{table}[h!]
\centering
\begin{tabular}{|c| c c c c c c c c|} \hline
   &   8    &   9   &   10   &  11   &  12   &   13    &    14   &   15 \\\hline
4  & 0.0544	& 0.0719  & 0.0888	& 0.1073 &	0.1264 &	0.1435 &	0.1606 &	0.1795  \\\hline
5  & 0.0440 &	0.0600	& 0.0752  & 0.0919 &	0.1092 &	0.1243 &	0.1393 &	0.1560 \\\hline
6  & 0.0184	& 0.0260	& 0.0325  &	0.0403 &	0.0485 &	0.0546 &	0.0607 &	0.0687  \\\hline
7  & 0.0100	& 0.0156	& 0.0199  &	0.0254 &	0.0312 &	0.0349 &	0.0385 &	0.0438 \\\hline
\end{tabular}
\caption{Consistency test for the S\&P data}
\label{tab4}
\end{table}
 
\section{Final remarks}
To sum up, the discrete time framework allows us to use the probabilistic relationship between the default probabilities and the credit transition matrix, to set up a constrained ill-posed, linear problem inverse problem for the CMT. Once the CTM matrix is available, one can use it to predict future cumulative default probabilities. In the test cases that we considered, this is rather satisfactory.

The procedure hinges on the fortunate fact that the default ratings are available yearly, and that discrete time modeling is  applicable in this case. It does not seem clear how to do the same in the continuous time framework. To begin with, the default data is provided in discrete time even though corporations can default at any time within a given year. Perhaps there is some measurable data that could be related to the absorption (default) time in continuous time modeling, which could then be used to develop a procedure similar to the one we proposed here.

To finish, we mention that once the matrix $\bQ^*$ is available, the quantity $T_{i,j}=\big(\bI - \bQ\big)^{-1}_{i.j}$ provides us with the average time that a corporation, initially rated in class $i,$ spends in a better risk rating class $j.$ Similarly, the $i$th component of the vector $\big(\bbI - \bQ\big)^{-1}\bu$ is the expected time to default of a corporation that had initial rating $i.$

\end{document}